\documentstyle[aps]{revtex}

\newcommand{\bq}{\begin{equation}}
\newcommand{\eq}{\end{equation}}
\newcommand{\bqn}{\begin{eqnarray}}
\newcommand{\eqn}{\end{eqnarray}}
\newcommand{\nb}{\nonumber}
\newcommand{\lb}{\label} 

\input epsf

\begin{document}
\draft

\twocolumn[\hsize\textwidth\columnwidth\hsize\csname
@twocolumnfalse\endcsname

\title{Critical collapse in tensor-multi-scalar and non-linear gravity 
theories: A universal class}
\author{Anzhong Wang}
\address{ Departamento de F\' {\i}sica Te\' orica,
Universidade do Estado do Rio de Janeiro, 
Rua S\~ ao Francisco Xavier 524, Maracan\~ a,
20550-013 Rio de Janeiro~--~RJ, Brazil }
 
\date{ November 23, 1998 }

\maketitle

\begin{abstract}

Critical collapse in tensor-multi-scalar gravity theories
is studied, and found that for any given target space all the 
theories conformally related belong to the same 
universal class. When only one scalar field is present,
the universality is extended to include a class of non-linear
gravity theories.
 
\end{abstract}


\vskip2pc]

 
 Studies of black hole formation from gravitational collapse of a 
 massless scalar field have revealed interesting non-linear phenomena 
 at the threshold of black hole formation \cite{Ch1993}. Specifically, 
 starting with spherical spacetimes,
 $$
 d\bar{s}^{2} = - \bar{\alpha}^{2}(t, r)dt^{2} 
 + \bar{a}^{2}(t, r)dr^{2}
 + r^{2}d\Omega^{2},
 $$
 where $d\Omega \equiv d\theta^{2} + \sin^{2}\theta d\varphi^{2}$, and
 $\{x^{\mu}\} = \{t, r, \theta, \varphi\}$ are the usual spherical 
 coordinates, Choptuik has investigated the Einstein-scalar field 
 equations and found the 
 following striking features: Consider a generic smooth one-parameter
 family of initial data, say, $S[p]$, such that for $p > p^{*}$, black 
 holes are formed, and for $p < p^{*}$ no black holes are formed.
 Then, (a) the critical solution is 
 {\em universal} with respect to the families of initial data considered, 
 and {\em periodic},
 $
 \bar{A}^{*}(\xi, \tau) = \bar{A}^{*}(\xi, \tau + \triangle),
 $
 where $\bar{A}^{*} = \{\bar{\alpha}^{*}, \bar{a}^{*}, 
 \bar{\phi}^{*}\}$, and 
 \bq
 \lb{eq2}
 \tau = \ln\left(\frac{t}{r_{0}}\right),\;\;\;
 \xi = \ln\left(\frac{r}{t}\right) - \xi_{0}(\tau), 
 \eq
 with $r_{0}$ being a dimensionful constant, $\bar{\phi}$ the massless
 scalar field, and $\xi_{0}(\tau)$ a 
 periodic otherwise arbitrary function with period $\triangle$. The 
 constant $\triangle$ is a {\em dimensionless universal} constant, which 
 has been numerically determined as $\triangle \approx 3.447$. 
 (b) Near the
 critical solution but with $p > p^{*}$,
 the black hole masses take a scaling form
 $
 M_{BH} = K (p - p^{*})^{\gamma},
 $
 where $K$ is a family-dependent constant, but $\gamma$ is another 
 {\em dimensionless universal} constant with $\gamma \approx 0.37$.
 The above results were soon confirmed by several independent studies both
 numerically \cite{Gu1994} and analytically \cite{Gu1995}, and extended to
 other matter fields, such as, axisymmetric gravitational waves \cite{AE1993},
 perfect fluids \cite{EC1994}, non-linear 
 $\sigma$-models \cite{HE1995,LC1996}, 
 Einstein-Yang-Millers fields \cite{CCB1996}, Einstein-Maxwell-scalar fields 
 \cite{GM1996}, and more recently Einstein-Skyrme
 fields \cite{BC1998}. 
 
 Now it is clear that in general the critical solution and the
 two dimensionless constants $\triangle$ and $\gamma$ 
 are universal only with respect to
 the same matter field, and  usually are matter-dependent.
 Moreover, the critical solutions can have discrete self-similarity (DSS)
 or continuous self-similarity (CSS), or none of them, depending on the matter 
 fields and regions of  the initial data spaces \cite{Gu1997}. 
 So far, in all the cases
 where the critical solution either has DSS or CSS, the formation of
 black holes always turns on with zero mass, the so-called Type II collapse,
 while in the cases in which the critical solution has neither DSS nor CSS,
 the formation always turns on with a mass gap, 
 the so-called Type I collapse. We shall first consider  type II 
 collapse, and then turn to type I collapse. 
 The universality of the critical solution and the exponent $\gamma$ now
 are well understood in terms of perturbations \cite{HKA1996}, 
 while the one of 
 $\triangle$ still remains somewhat of a mystery. The former is 
 closely related to the fact that the critical solution has only one unstable
 mode. This property now is considered as the main criterion for a solution to
 be critical \cite{Gu1997}.
 
 In this Letter, we shall consider the critical collapse in 
 tensor-multi-scalar (TMS)
 and non-linear (NI) gravity theories, and pay special attention on 
 universality of the collapse, that is, systems that have the same critical 
 behavior near their critical points. This is particularly interesting 
 in gravitational collapse,  as
 the field equations involved are 
 usually very complicated \cite{Jo1993}, and if we could find a 
 universal class, then we only need to 
 study the simple ones of the class in order to know
 the properties of others that usually 
 are too complicated to be studied either 
 analytically or numerically.  
 Yet, the  TMS and
 NI gravity theories are the most 
 promising alternatives to Einstein's
 theory, and are well motivated from superstrings and quantum gravity 
 \cite{MS1994}. It has been shown that the  TMS
 theories can equally pass all the known observational and experimental 
 constraints as GR does, while can be significantly different from GR in 
 the strong field regime. 
 These differences are within the scope of the
 undertaking experiments, such as, LIGA and VIRGA \cite{DF1992}. NI
 gravity theories are usually connected with quantum corrections and are
 essentially also related to the strong field regime \cite{St1980}. 
 Since critical phenomena are strong field ones, it would be
 very interesting to study critical collapse
  both in  TMS and NI theories. 
 
 So far, only two universal classes have been found, 
 one is the massless scalar field with any potential \cite{Ch1993}, 
 and the other is a fluid with the equation of state, $p = k \rho + f(\rho)$ 
 for any smooth function $f(\rho)$ \cite{HKA1996}. However, they are all
 within the framework of GR. In this Letter we shall provide the third class 
 but in terms of different theories of gravity.
 
 The field equations for  the TMS theories are given by 
 \cite{DF1992}
 \bqn
 \lb{eq3}
 \bar{G}[\bar{g}]_{\mu\nu} &=& 8\pi G \left\{ \gamma_{ab}(\bar{\phi})
 [\bar{\phi}^{a}_{,\mu}\bar{\phi}^{b}_{,\nu} - \frac{1}{2}
 \bar{g}_{\mu\nu} \bar{\phi}^{a}_{,\alpha}\bar{\phi}^{b,\alpha}]
 \right.\nb\\
 & & \left. - B(\bar{\phi}) \bar{g}_{\mu\nu}\right\},\\
 \lb{eq4}
 \bar{\Box}\bar{\phi}^{a} &=& B^{a}(\bar{\phi})
  - \gamma^{a}_{bc}(\bar{\phi})\bar{\phi}^{b,\alpha}
  \bar{\phi}^{c}_{,\alpha},
  \eqn
  where $\bar{\phi}^{a}\; (a = 1, 2, ..., N)$ denote the $N$ scalar fields,
  $B(\bar{\phi})$ the potential, $B_{a}(\bar{\phi}^{c}) \equiv
  \partial B(\bar{\phi}^{c})/\partial \bar{\phi}^{a}$,
 and $\gamma_{ab}(\bar{\phi})$ is the
  metric in the target space, while $\gamma^{a}_{bc}(\bar{\phi})$ are
  the corresponding Christoffel symbols. From now on we shall use 
  quantities with bars to denote the ones defined in the Einstein frame, 
  and quantities without bars to denote the ones defined in the
  Jordan frame which will be considered as
   the physical one \cite{MS1994,DF1992}. In the
  following, we shall first consider the cases $N = 1, \; 2$, and then 
  generalize the obtaining results to the case for any $N$.
 
 When only one scalar field is present, Eqs.(\ref{eq3}) and (\ref{eq4})
 reduce to the corresponding Einstein-scalar equations
 considered in \cite{Ch1993}. 
 making the conformal transformation
 \bqn
 \lb{eq5}
 \bar{g}_{\mu\nu} &=& f(\phi) g_{\mu\nu},\nb\\
 \bar{\phi}(\phi) &=& \int^{\phi}_{\phi_{0}}
 \left[\frac{3f'(x)^{2}}{2 f(x)^{2}}
 + \frac{h(x)}{f(x)}\right]^{1/2}dx,
 \eqn
 we find that 
 $
 S[\bar{g},\bar{\phi}] = \int(- \bar{g})^{1/2}(\bar{R} - 
 \bar{\phi}_{,\alpha}\bar{\phi}^{,\alpha}) dx^{4} 
  =
 \int({- {g}})^{1/2}[f(\phi){R} - 
 h(\phi){\phi}_{,\alpha}{\phi}^{,\alpha}] dx^{4},
 $
 which shows that in the Jordan frame the action represents the general
 Bergman-Wagoner-Nordtvedt (BWN) tensor-scalar theories 
 \cite{MS1994,DF1992}, where $f(\phi)$ and
 $h(\phi)$ are {\em strictly positive and non-singular}, and a prime denotes
 the ordinary differentiation. The BWN theories include several interesting
 cases, such as, $f(\phi) = 1 - \lambda \phi^{2},\; h(\phi)  
 = 1$. This is the only case where the coupling constant $\lambda$ is  
 dimensionless. In this case critical collapse has been 
 studied numerically  \cite{Ch1992} and 
 found that the two constants $\triangle$ and $\gamma$ are weakly dependent 
 on $\lambda$, while by dimensional analysis
 some kind of dependence has been expected 
 \cite{HE1995,Gu1997}. 
 Another interesting case is $f(\phi) = \phi,\;
 h(\phi) = \omega/\phi$, which corresponds to the Brans-Dicke (BD) theory,
 where $\omega$ is the BD coupling constant.
 In this case critical collapse
  has been also studied numerically \cite{CS1996}, and found 
 that $\triangle$ depends on $\omega$ weakly while 
 $\gamma$ strongly.  We shall show  that all the
 BWN theories belong to the same universal class and 
 have the same $\triangle$ and $\gamma$. 
 
 To show the claim, let us first prove that
 the solution $A^{*} = \{g_{\mu\nu}^{*}, \phi^{*}\}$, which is obtained
 through Eq.(\ref{eq5}) from the critical solution $\bar{A}^{*}$ found
 in \cite{Ch1993},
 is a critical solution in the BWN theories. First, since $f(\phi)$ is
 non-singular, the solution $A^{*}$ has the same singular
 behavior as $\bar{A}^{*}$ according to Penrose's 
 theorems \cite{Pe1965}. In particular, if $\bar{A}^{*}$ is no singular
 on the axis $r = 0$ for any $t < 0$, so is $A^{*}$.
  Moreover, $\bar{\phi}$ is
 a monotonically increasing function of $\phi$
 [otherwise, we can restrict the mapping to a specific
 region, so that it is one-to-one], so the inverse $\phi = 
 \phi(\bar{\phi})$ exists and $\phi^{*}$ has the same period as 
 $\bar{\phi}^{*}$. Thus, 
 if $\bar{A}^{*}$ has period $\triangle$ 
 so does $A^{*}$. Second, if the perturbations of the critical
 solution $\bar{A}^{*}$ are given by \cite{Gu1995}
 \bq
 \lb{eq6}
 \delta \bar{Z} = \sum_{i = 1}^{\infty}C_{i}(p)e^{\lambda_{i}\tau}
 \delta_{i}\bar{Z}(\xi,\tau),
 \eq
 where $\delta_{i}\bar{Z}(\xi,\tau)$ are periodic functions with period
 $\triangle$, then from Eq.(\ref{eq5}) 
 it can be shown that the perturbations
 of the solution $A^{*}$ to first order take the same form 
 only with $\bar{Z}$ being replaced
 by $Z$ in Eq.(\ref{eq6}), where $\bar{Z}
  \equiv \{\bar{\alpha}, \bar{a}, \bar{\phi}_{,\xi},
 \bar{\phi}_{,\tau}\}$ \cite{Gu1995}. 
 Since the equations satisfied
 by $\delta_{i}{Z}$ are related with the ones satisfied by 
 $\delta_{i}\bar{Z}$ through the regular transformation (\ref{eq5}), the
 perturbations have the same spectrum in both cases. Thus, if
 $\delta_{i}\bar{Z}$ has only one unstable mode, say, $\lambda_{1}$, so
 does $\delta_{i}Z$. Therefore, if $\bar{A}^{*}$ is a critical solution
 of the Einstein-scalar field equations, $A^{*}$ is a critical solution
 of the BWN field equations. 
 
 Assuming that the linear perturbations are valid until the moment $\tau
 = \tau_{0}$, we have
 \bq
 \lb{eq7}
 Z_{p}(r) \equiv Z(\xi, \tau_{0}) \nb\\
 = Z^{*}(\ln \frac{r}{r_{p}},\tau_{0})
 + \epsilon \delta_{1}Z(\ln \frac{r}{r_{p}}, \tau_{0}),
 \eq
 where $r_{p} \equiv r_{0}e^{\tau_{0}+\xi_{0}(\tau_{0})}$, and $
 \epsilon = (p - p^{*})e^{\lambda_{1}\tau_{0}}\partial C_{1}(p^{*})/
 \partial p$ is very small so that the perturbations are valid. Note that
 in writing the above expressions, we have dropped all the stable modes and
 expanded $C_{1}(p)$ to its first order by using the fact that $C_{1}(p^{*})
 = 0$, because when $p = p^{*}$ we shall obtain exactly the critical
 solution. To show the explicit dependence on $r_{p}$, 
 we have also replaced $\xi$ by $\ln (r/r_{p})$.
 Now we need to use Eq.(\ref{eq7}) 
 as the new initial data to consider the non-linear evolution of the 
 BWN field equations, which in the Jordan frame read 
 \bqn
 \lb{eq8}
 X_{\pm,\xi} &=&\frac{C_{\pm} \pm e^{\xi+\xi_{0}}g (X_{\pm,\tau} + 
 X_{\pm}W_{,\tau}/W)}{1 \pm ( 1 + \xi_{0}')e^{\xi+\xi_{0}}g},\nb\\
 a_{,\xi} &=& \frac{1}{f^{1/2}}\left\{C_{1} - 
 \frac{a^{2}f'X}{({8\pi G} f)^{1/2}}
 \right\},\nb\\
 a_{,\tau} &=& \frac{1}{f^{1/2}}\left\{(1+\xi'_{0})C_{1}
 + \frac{C_{2}}{ge^{\xi+\xi_{0}}} \right.\nb\\
 &&\left. - \frac{\alpha f'}{({8\pi G}f)^{1/2}}
 \left[\frac{\alpha Y}{e^{\xi + \xi_{0}}} + (1 + \xi_{0}')aX\right]
 \right\},\nb\\
 g_{,\xi} &=& (1 - fa^{2})g,
 \eqn
 where 
 $g \equiv \alpha/a,\; 
 X \equiv (2\pi G)^{1/2}\phi_{,r}/a,\;
 Y \equiv (2\pi G)^{1/2}\phi_{,t}/\alpha, \; 
 W \equiv f^{-1/2}d\bar{\phi}/d\phi,\; 
 X_{\pm} \equiv X \pm Y$,
 and $C_{1},\; C_{2}$, and $C_{\pm}$ are functions of 
 $a,\; f,\; W$ and $X_{\pm}$
 only. Now the key observation is that 
 the initial data Eq.(\ref{eq7}) depend
 on $r_{p}$ only through the form $r/r_{p}$ and $t_{0}/r_{p}$, while
 the field equations (\ref{eq8}) are invariant under the translation of 
 $\tau$. Then, the entire solution must take the
 scale-invariant form \cite{HE1995}
 \bq
 \lb{eq9}
 Z(\xi,\tau) = Z(\ln\frac{r}{r_{p}}, \frac{t}{r_{p}}).
 \eq
 On the other hand, we know that the mass of black holes has dimension 
 length, while in the above solution the only quantity that has such 
 a dimension is $r_{p}$. Therefore, we must have
 \bq
 \lb{eq10}
 M_{BH} \propto r_{p} = r_{0}e^{\xi_{0} + \tau_{1}}(p - p^{*})^{\gamma},
 \eq
 where $\gamma = -\lambda_{1}^{-1}, \; \tau_{1} = \epsilon^{-1}\gamma\ln
 [\partial C_{1}(p^{*})/\partial p]$. 
 The above analysis shows that {\em the
 gravitational collapse of the BWN scalar field in the BWN
 theories exhibits the same critical behavior as that of a massless scalar 
 field in Einstein's theory. In particular, they have the same
 $\triangle$ and $\gamma$}. Therefore, they belong to the same
 universal class.
 
 Now let us turn to the NI theories given by the action 
 \bq
 \lb{eq11}
 S_{non.}[g, R] = \int 
 d^{4}x (-g)^{1/2}f(R), 
 \eq
 where $f(R)$ is an analytic function of the
 Ricci scalar $R$. In this case the collapse of the gravitational field
 also exhibits the same critical behavior as the massless scalar field
 in Einstein's theory studied in \cite{Ch1993}. This is because by the
 conformal transformation
 $
 e^{\lambda_{0} \bar{\phi}} = f'(R),\;\;\; \bar{g}_{\mu\nu} 
 = e^{\lambda_{0} \bar{\phi}}g_{\mu\nu},
 $
 we find that
 $
 S_{non.}[g,R] = \int 
 d^{4}x (-g)^{1/2}f(R) = \int
 d^{4}x(-\bar{g})^{1/2}[\bar{R}  - 
 \bar{\phi}^{,\alpha}\bar{\phi}_{,\alpha}
 - 2V(\bar{\phi})],
 $
 where 
 $V(\bar{\phi}) = e^{-\lambda_{0} \bar{\phi}}
 [R(\bar{\phi}) - e^{-\lambda_{0}\bar{\phi}}f(R(\bar{\phi}))]/2,
 $
 and
 $\lambda_{0} = (2/3)^{1/2}$. That is, the non-linear gravity theories
 given by the action (\ref{eq11}) is related to the Einstein-scalar
 theory by a conformal transformation, where the scalar field $\bar{\phi}$
 usually has a non-vanishing potential $V(\bar{\phi})$. As shown 
 numerically in \cite{Ch1993} and analytically in \cite{HKA1996}, the
 potential  does not affect the (type II) critical behavior,
 so for any given potential the collapse will exhibit the same critical
 phenomena as the one with vanishing potential. Then, using the same
 conformal transformation
 arguments as given in the BWN case, we can show that 
 the critical collapse of the gravitational field in the NI 
 theories exhibits the same critical
 behavior as a scalar field with a potential in Einstein's theory.
 Therefore, {\em the gravitational collapse of a pure gravitational field
 in the NI theories given by Eq.(\ref{eq11})
 and the one of the BWN gravitational scalar field in the BWN theories
 all belong to the same universal class as that of a scalar field 
 (with or without potential) in Einstein's theory}.

 When two scalar field are present, a particular case was studied
 in \cite{HE1995,LC1996}, where the functions $\gamma_{ab}(\bar{\phi}^{c})$
 was chosen as $\gamma_{ab} = \lambda^{-1}
  \; diag. \{1, g(\bar{\phi}^{1})\}$, with
 $g(\bar{\phi}^{1}) = e^{-\bar{\phi}^{1}},\; \sin^{2}2\bar{\phi}^{1}$,
 and $\lambda$ being a coupling constant. 
 When $g = e^{-\bar{\phi}^{1}}$, the non-linear $\sigma-$model is equivalent
 to the BD theory coupled with a massless scalar field $\bar{\phi}^{2}$
 with $\lambda = 8 \kappa = (\omega + 3/2)^{-1} > 0$, which has been studied
 in details in \cite{LC1996}, while when $g = \sin^{2}2\bar{\phi}^{1}$, it
 is the case studied in \cite{HE1995} for $\kappa = \lambda/8 < 0$,
 where $\kappa$ is a constant. As noticed
 in \cite{HE1995}, in the latter case there is no correspondence between the
 NI $\sigma-$model and the BD theory. For more details, we refer
 readers to \cite{HW1998}. It has been shown 
 \cite{HE1995,LC1996} that critical (Type II)
 behavior exists for all the range of $\kappa$, although the properties
 of the collapse crucially depend on $\kappa$. 
  In particular,  For $\kappa > 1/12$, 
 a family of critical solutions characterized by $\kappa$ exists, 
 and all these solutions have CSS. The exponent $\gamma$
 strongly depend on $\kappa$ in this range. 
 For $0 < \kappa < 1/12$,  the critical solution found in 
 \cite{Ch1993} seems to be the only one, 
 and both $\gamma$ and $\triangle$
 are now weakly dependent on $\kappa$. For $\kappa < - 0.28$ further 
 bifurcations are present. In \cite{HE1995,LC1996}, 
 the problem was studied in the
 Einstein frame. An open question is that 
 do the results still hold in the BD
 theory (or in the Jordan frame) \cite{Li1995}?
 as it is well-known that
 the mass of black holes are not well defined in the BD theory 
 \cite{MS1994}.
 In the following, we shall use the conformal transformation arguments to
 show that this indeed is the case. 
 
 The conformal transformation from the
 Einstein theory to the BD theory is given by $\bar{g}
 _{\mu\nu} = f(\bar{\phi}^{1}) g_{\mu\nu}$ with $f(\bar{\phi}^{1}) = 
 \bar{\phi}^{1}$. However, to have 
 the results applicable as much as possible, 
 we shall take $f(\bar{\phi}^{1})$ as a non-singular otherwise arbitrary
 function of $\bar{\phi}^{1}$. Moreover, we will not consider 
 only the case where $\kappa > 0$, but also the case where
 $\kappa < 0$, although in the latter case no known theories 
 are related. Since the conformal factor $f(\bar{\phi}^{1})$ is 
 non-singular, we can see that the metric in the Jordan frame
 will have the same singular behavior as that in the Einstein
 frame. Yet, if the solution $\bar{Z}^{*}$
 is a critical solution in the Einstein frame, the solution $Z^{*}$ is
 a critical solution in the Jordan frame, because the equations for
 the perturbations $\delta \bar{Z}$ are related to the ones for $\delta
 Z$ by a regular conformal transformation, and consequently they have the
 same spectrum of the unstable modes in both cases, where $\bar{Z}$  
 represents a set of variables of fields such that the complete field 
 equations can be written in the first-order form 
 $
 \bar{F}(\bar{Z}, \bar{Z}_{,\tau},\bar{Z}_{,\xi}) = 0. 
 $
 Moreover, if this equation is scale-invariant under 
 the translation of 
 $\tau$, so is the equation 
 $
 {F}({Z}, {Z}_{,\tau}, {Z}_{,\xi}) = 0
 $ for $Z$. Then, using the scale-invariant arguments
 given previously we can see that the entire solutions must take
 the scale-invariant form Eq.(\ref{eq9}),
 and as a result, the mass of black holes for the 
 supercritical case will take the same scaling form in the 
 both theories. 
 
 Clearly, the above analysis does not depend on the particular choice
 of $\gamma_{ab}(\bar{\phi}^{1}, \bar{\phi}^{2})$ and 
 $f(\bar{\phi}^{1}, \bar{\phi}^{2})$. Therefore, for any given theory,
 that is, for any given target space 
 $\gamma_{ab}(\bar{\phi}^{1}, \bar{\phi}^{2})$, all the
 theories related to it by the conformal 
 transformation $\bar{g}_{\mu\nu}
 = f(\bar{\phi}^{1}, \bar{\phi}^{2}) g_{\mu\nu}$ belong to
  the same universal class. Moreover, if the
 critical solution $\bar{Z}^{*}$ has CSS (DSS), the solution $Z^{*}$
 has CSS (DSS), too. In the DSS case, a ``wiggle" overlaid 
 on the scaling law was found \cite{Gu1995}. 
 It is not difficult to see that it will persist after 
 the conformal transformation. 
 
 It is not difficult to be convinced  that 
 the above conclusions do not depend on how many scalar fields
 $\bar{\phi}^{a}$ involved and the explicit form of $f(\bar{\phi}^{a})$,
 as long as $f(\bar{\phi}^{a})$ is {\em positive and 
 non-singular}. Also, they do not depend on the {\em symmetries}, such as,
 spherical, axial, etc., of the spacetimes. 
 
 For type I collapse, using the same conformal transformation arguments 
 as given in the type II case,
 it can be shown that if $\bar{Z}^{*}$ is a type I critical solution
 in the Einstein frame, $Z^{*}$  
 is also a type I critical solution in the Jordan frame. 
 This is because in this case the perturbations $\delta \bar{Z}$ will 
 take the form \cite{Gu1997}
 \bq
 \lb{eq14}
 \delta \bar{Z}(t, r) = \frac{\partial C_{1}(p^{*})}{\partial p}
 e^{\lambda_{1} t}\delta_{1}\bar{Z}(t, r),
 \eq
 in the Einstein frame. Clearly after the conformal transformation, 
 the perturbations $\delta Z$ to the first order 
 will take the same form. Since the equations satisfied 
 by $\delta \bar{Z}$ are conformally related to the ones
 satisfied by $\delta {Z}$,  they must have the
 same spectrum of the perturbation modes. Thus, if $\bar{Z}^{*}$ is
 a critical solution in the Einstein's frame, $Z^{*}$ is a critical
 solution in the Jordan frame. On the other hand,
 Eq.(\ref{eq14}) is different
 from Eq.(\ref{eq6}) in that in the former 
 the perturbations are proportional
 to $e^{\lambda_{1}t}$, while in the latter they are proportional
 to $(- t)^{\lambda_{1}}$. It is this difference that determines the 
 types of the collapse. Since the conformal transformation does not change
 the forms of perturbations,
 it is concluded that {\em it does not change the types of collapse, 
 and that   all the theories that are
 related by a conformal transformation belong to the same universal
 class for type I collapse, too}.  
 
 In conclusion, we have shown that {\em all the theories that are 
 related by a conformal transformation belong to the same universal 
 class, irrespective of the spacetime symmetries.
 The conformal transformation preserves
 the types, Type I, or Type II, of the collapse and the 
 symmetries, CSS or DSS, of the critical solutions}. As a result,
 all these theories are distinguishable near their critical points.


The author would like to express his gratitude to M.W. Choptuik
and E.W. Hirschmann for valuable suggestions and discussions.  
Part of the work was done while he was a visitor at 
the Center for Relativity, the University of Texas, Austin. He would like
to thank  the Center for hospitality. The financial 
assistance from CNPq, FAPERJ, and UERJ, is gratefully acknowledged.

\end{document}